\documentclass[a4paper,preprintnumbers,floatfix,superscriptaddress,aps,prl,twocolumn,showpacs,notitlepage,longbibliography]{revtex4-1}
\pdfoutput=1
\usepackage[utf8]{inputenc}
\usepackage{graphicx}
\usepackage{times,bbm,amsmath,amssymb,braket}
\usepackage{hyperref}
\usepackage{bm}

\DeclareMathOperator{\Tr}{Tr}

\begin{document}

\title{Quantum violation of local causality in urban network with hybrid photonic technologies}
\date{\today}

\author{Gonzalo Carvacho}
\thanks{These two authors contributed equally}
\affiliation{Dipartimento di Fisica - Sapienza Universit\`{a} di Roma, P.le Aldo Moro 5, I-00185 Roma, Italy}

\author{Emanuele Roccia}
\thanks{These two authors contributed equally}
\affiliation{Dipartimento di Fisica - Sapienza Universit\`{a} di Roma, P.le Aldo Moro 5, I-00185 Roma, Italy}

\author{Mauro Valeri}
\affiliation{Dipartimento di Fisica - Sapienza Universit\`{a} di Roma, P.le Aldo Moro 5, I-00185 Roma, Italy}

\author{Francesco Basso Basset}
\affiliation{Dipartimento di Fisica - Sapienza Universit\`{a} di Roma, P.le Aldo Moro 5, I-00185 Roma, Italy}

\author{Davide Poderini}
\affiliation{Dipartimento di Fisica - Sapienza Universit\`{a} di Roma, P.le Aldo Moro 5, I-00185 Roma, Italy}

\author{Claudio Pardo}
\affiliation{Dipartimento di Fisica - Sapienza Universit\`{a} di Roma, P.le Aldo Moro 5, I-00185 Roma, Italy}

\author{Emanuele Polino}
\affiliation{Dipartimento di Fisica - Sapienza Universit\`{a} di Roma, P.le Aldo Moro 5, I-00185 Roma, Italy}

\author{Lorenzo Carosini}
\affiliation{Dipartimento di Fisica - Sapienza Universit\`{a} di Roma, P.le Aldo Moro 5, I-00185 Roma, Italy}

\author{Michele B. Rota}
\affiliation{Dipartimento di Fisica - Sapienza Universit\`{a} di Roma, P.le Aldo Moro 5, I-00185 Roma, Italy}

\author{Julia Neuwirth}
\affiliation{Dipartimento di Fisica - Sapienza Universit\`{a} di Roma, P.le Aldo Moro 5, I-00185 Roma, Italy}

\author{Saimon F. Covre da Silva}
\affiliation{Institute of Semiconductor and Solid State Physics, Johannes Kepler University, 4040 Linz, Austria }

\author{Armando Rastelli}
\affiliation{Institute of Semiconductor and Solid State Physics, Johannes Kepler University, 4040 Linz, Austria }

\author{Nicol\`o Spagnolo}
\affiliation{Dipartimento di Fisica - Sapienza Universit\`{a} di Roma, P.le Aldo Moro 5, I-00185 Roma, Italy}

\author{Rafael Chaves}
\affiliation{International Institute of Physics and School of Science and Technology, Federal University of Rio Grande do Norte, 59078-970, P. O. Box 1613, Natal, Brazil}

\author{Rinaldo Trotta}
\email{rinaldo.trotta@uniroma1.it}
\affiliation{Dipartimento di Fisica - Sapienza Universit\`{a} di Roma, P.le Aldo Moro 5, I-00185 Roma, Italy}

\author{Fabio Sciarrino}
\email{fabio.sciarrino@uniroma1.it}
\affiliation{Dipartimento di Fisica - Sapienza Universit\`{a} di Roma, P.le Aldo Moro 5, I-00185 Roma, Italy}

\begin{abstract}
Quantum networks play a crucial role for distributed quantum information processing, enabling the establishment of entanglement and quantum communication among distant nodes. Fundamentally, networks with independent sources allow for new forms of nonlocality, beyond the paradigmatic Bell's theorem. Here we implement the simplest of such networks --the bilocality scenario-- in an urban network connecting different buildings with a fully scalable and hybrid approach. Two independent sources using different technologies, respectively a quantum dot and a nonlinear crystal, are used to share photonic entangled state among three nodes connected through a 270 m free-space channel and fiber links. By violating a suitable non-linear Bell inequality, we demonstrate the nonlocal behaviour of the correlations among the nodes of the network. Our results pave the way towards the realization of more complex networks and the implementation of quantum communication protocols in an urban environment, leveraging on the capabilities of hybrid photonic technologies.
\end{abstract}

\maketitle

\emph{Introduction -- }
In the last decade, several breakthroughs on quantum communication have been reported, especially those regarding the experimental realization of quantum networks \cite{gisin2007quantum,lo2014secure,diamanti2016practical,wehner2018quantum}.
Quantum key distribution on fiber networks have demonstrated the possibility to securely connect distances greater than $400$ km \cite{chen2020sending,boaron2018secure,yin2016measurement} and the successful launch of a satellite allowed the first quantum network covering record distances over $4,600$ km, integrating space-to-ground and optical fibers communication \cite{chen2021integrated}. At the basis of many of these quantum communication protocols is the phenomenon of Bell nonlocality \cite{bell1964einstein,brunner2014bell,scarani2019bell}, arguably the most radical departure between classical and quantum descriptions of nature. Besides its profound foundational implications, generating nonlocal correlations has become of crucial importance for a variety of quantum technologies, ranging from distributed computing \cite{buhrman1}, quantum cryptography \cite{crypt1,crypt2,crypt3,crypt4,crypt5,scarani2009security,xu2020secure} and quantum key distribution \cite{barret1,acin1} to randomness generation \cite{pironio1,liu1,agresti1} and self-testing \cite{vsupic2020self,agresti2}. 

Despite the apparent simplicity of Bell's theorem ~\cite{bell1964einstein}, it tooks over fifty years for the first loophole-free violation of a Bell inequality \cite{Giustina,Shalm,Hensen}. This has been achieved considering the simplest Bell scenario where a single source distributes entangled pairs among two distant nodes. Within this context, nonlocal correlations have been obtained using free-space links \cite{peng2005experimental,ma2012quantum,yin2012quantum,basset2021quantum}, fiber-based links \cite{wengerowsky2019entanglement,ikuta2018four,dynes2009efficient,inagaki2013entanglement,cozzolino2019air,Schimpf2021}, and satellite-based communications \cite{yin2017satellite,bedington2017progress,ren2017ground,yin2020entanglement}.
Moving beyond the paradigmatic Bell scenario, it has been realized that nonlocality can also arise in more complex networks, where the correlations between the distant nodes are mediated by a number of independent sources and in a variety of topologies. Motivated by a causality perspective \cite{pearl2009causality} showing, in particular that Bell's theorem can be seen as a particular case of a causal inference problem \cite{wood2015lesson,chaves2015unifying}, quantum networks of growing size and complexity have been attracting a growing theoretical interest \cite{branciard2010characterizing,branciard2012bilocal,tavakoli2021bilocal,renou2021quantum,chaves2015information,tavakoli1,chaves2016polynomial,renou1,fritz2016beyond,henson2014theory,wolfe2019inflation,gisin2019entanglement,renou2019genuine,renou2021quantum, tavakoli2021bell}.

In spite of its clear foundational and technological relevance, however, the experimental investigation  of these new forms of nonlocality~\cite{carvacho2017experimental,saunders2017experimental,andreoli2017experimental,sun2019experimental,poderini2020experimental,chaves2018quantum,polino2019device,carvacho2019perspective,ringbauer2016experimental,agresti2021experimental} have been hampered by difficulties arising from quantum networks. 
Indeed, in order to realize large scale quantum networks and exploit them for practical tasks \cite{wehner2018quantum,kimble2008quantum}, it is crucial to extend their implementation to urban scale scenarios. With that aim, two requirements have to be satisfied. First, the experimental apparatus should be scalable, such that connecting an increasing number of distant nodes is within technological reach. Second, the possibility to interface or merge different quantum technologies and types of communication links.

In this work, we take a significant step in this direction, by experimentally realizing a bilocal network \cite{branciard2010characterizing,carvacho2017experimental,andreoli2017experimental,branciard2012bilocal, sun2019experimental,poderini2020experimental}, a scenario akin to the paradigmatic entanglement swapping protocol \cite{zukowski1993event,zukowski1995entangling}. Importantly, differently from previous experiments, we generate nonlocal correlations in this network through a scalable and hybrid photonic platform composed of two photonic sources distributing photons among three nodes. In contrast with implementations that rely on entangled measurements, a demanding task in linear optics \cite{lutkenhaus1999bell,vaidman1999methods,BassoBasset2021}, we exploit separable measurements only, allowing the scalability of our approach to networks of increasing size.
Notably, the independent sources of quantum entanglement employ two radically different technologies: Spontaneous Parametric Down Conversion (SPDC) and a quantum dot (QD). SPDC represents the most used process for the generation of polarization entangled photon pairs \cite{eisaman2011invited}, providing high quality entangled states which can be generated in a compact and cost-effective system. In turn, QD is one of the most promising platforms for realizing deterministic photon sources \cite{huber2018strain}. Complementing the truly hybrid nature of our experiment, we  simultaneously employ a fiber-based as well as free-space communication link, arguably the main kinds of communication channels to be used in future urban quantum networks. In particular, the free-space link is a 270 m long connection between two buildings at the campus of Sapienza University of Rome. By violating a Bell inequality suited for the bilocal scenario, we demonstrate the nonlocal nature of the correlations generated among the nodes of the network. This is achieved in the so-called device independent paradigm \cite{pironio2016focus,vazirani2019fully,scarani2012device,poderini2021ab,arnon2018practical,ahrens2012experimental}, that is, without assuming any knowledge of the inner workings of the sources, measurement stations and other devices. In this way, we provide a reliable and versatile platform for a urban quantum network, feasible to be extended to complex scenarios with larger number of nodes, sources and covering larger distances. 

\begin{figure}[hb!]
\centering
\includegraphics[width=0.99\columnwidth]{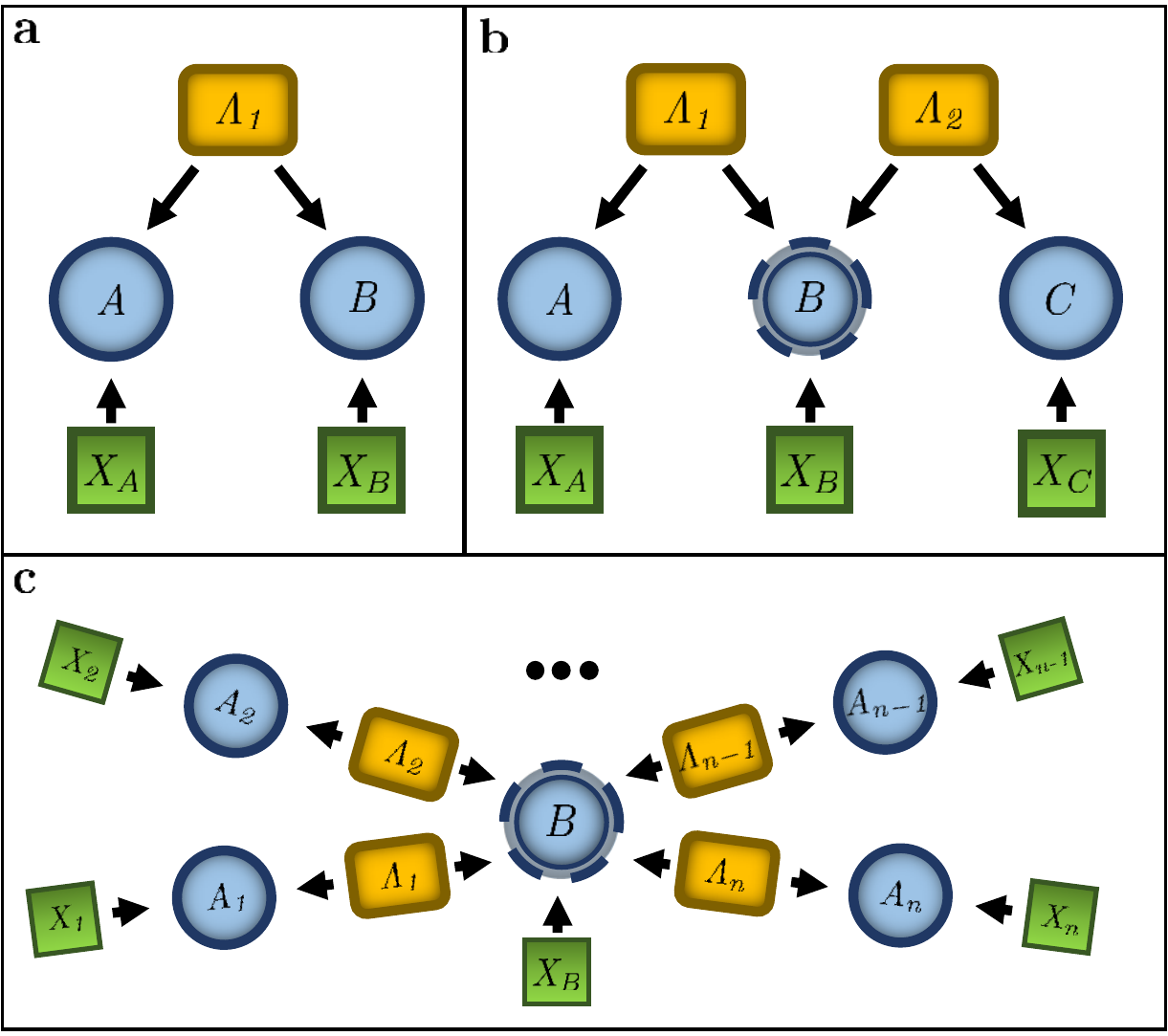}
\caption{ \textbf{Directed acyclic graph representation of different casual structures.} Examples of network where $n$-nodes are connected to a central one by means of intermediate nodes, i.e., the star-shape network (SSN)\cite{poderini2020experimental}. \textbf{a)} Standard Bell scheme. \textbf{b)} Bilocal scenario. \textbf{c)} General SSN scheme. Independent sources of correlations ($\Lambda_1, \dots ,\Lambda_n$) connect peripheral nodes ($\mathrm{A}_1,\dots,\mathrm{A}_n$) to a central one (B). The measurement performed by the nodes are also influenced by their measurement choices ($X_\mathrm{B},X_{1},\dots,X_{n}$). In particular, the central node B consists of different measurement setups which are influenced by different sources, $\Lambda_1, \ldots, \Lambda_n$, and the same measurement choice $X_\mathrm{B}$.  a) and b) are particular cases of SSN having $n=1$ and $n=2$, respectively.}
\label{fig:DAGnew}
\end{figure}

\begin{figure*}[t!]
\centering
\includegraphics[width=\textwidth]{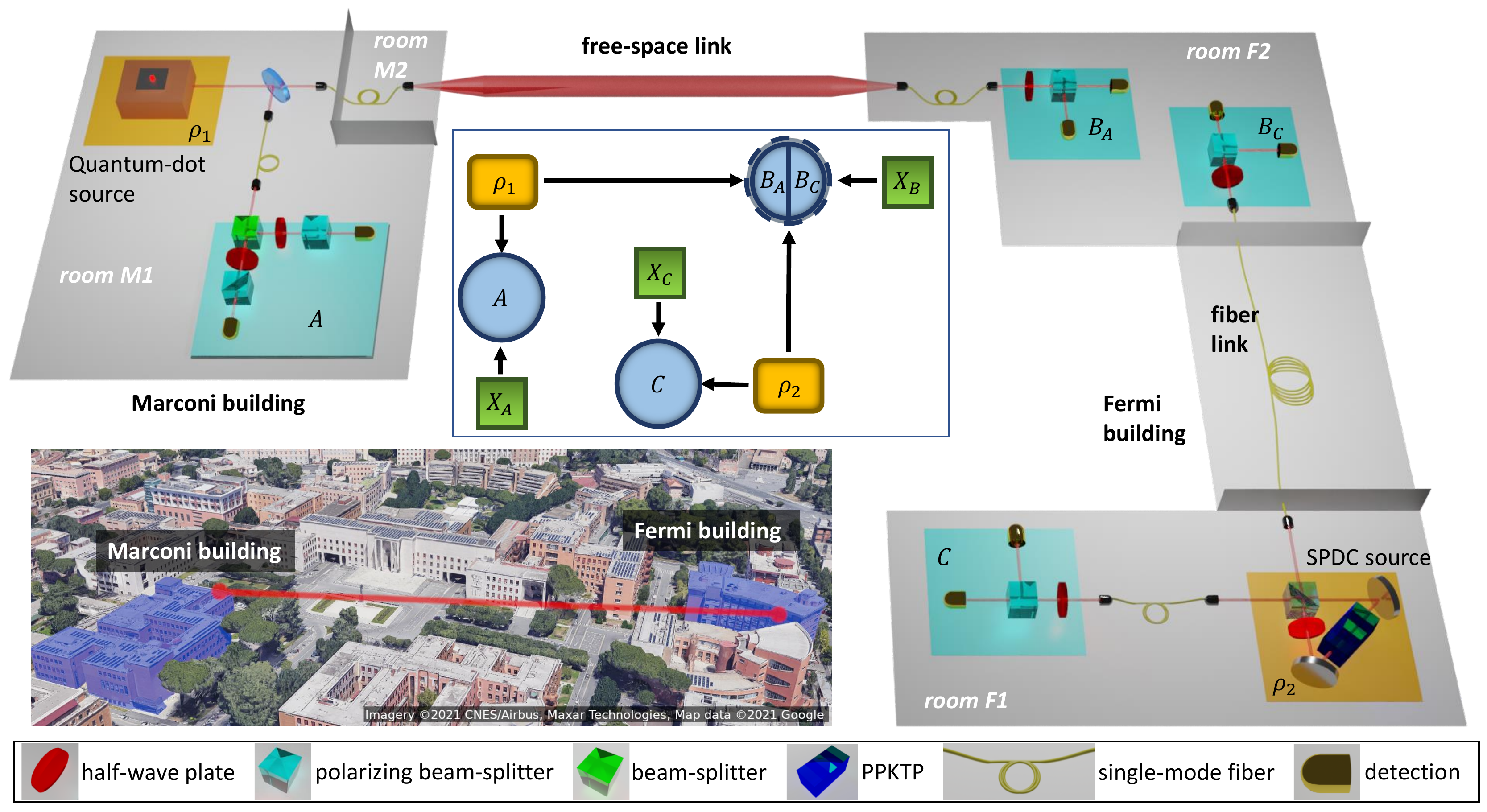}
\caption{ \textbf{Experimental implementation of the quantum network.} To realize the bilocal scenario, multiple laboratories located in different rooms and buildings were used. In particular, two sources of polarization-entangled photon pairs are realized via quantum dot device and SPDC in a Sagnac interferometer. They are placed inside the laboratories of two different buildings, respectively, the Marconi and the Fermi buildings. The entanglement is distributed from such laboratories to a central one --- place in the Fermi building --- by using a free space channel and a fiber link. A dedicated stabilization system was employed to use the free-space link (not shown in the figure). The corresponding bilocal scheme in DAG representation is reported in the middle: according to Fig.\ref{fig:DAGnew}b, two independent sources of correlations ($\rho_1, \rho_2$) connect two external nodes (A,C) to a central one (B). The two measurement setups influenced by the different sources are indicated as $\mathrm{B_A}$ and $\mathrm{B_C}$. 
}
\label{fig:DAGbilocale}
\end{figure*}


\emph{Bilocal Scenario -- } In its essence, Bell's theorem shows that the correlations obtained by measurements on distant parties of an entangled system cannot be explained by classical notions of cause and effect. In practice, we impose a given causal structure to our experiment and test whether the constraints arising from a classical description of it, the so called Bell inequalities, can be violated. The paradigmatic causal structure in Bell's theorem is shown in Fig.~\ref{fig:DAGnew}a, represented by a directed acyclic graph (DAG), where each node defines a variable of relevance for the experiment and the directed edges encode their causal relations. In Bell's DAG, two distant parties are connected by a single source, classically described by a hidden variable $\Lambda$, originating the correlations between the measurement outcomes $A$ and $B$ given that the parties measure observables parametrized by $X_A$ and $X_B$, respectively.

Moving beyond this simple causal structure, it has been realized that networks of different topologies involving an increasing number of nodes and independent sources can lead to new forms of nonlocality  \cite{branciard2012bilocal,chaves2015information,tavakoli1,chaves2016polynomial,renou1,fritz2016beyond,henson2014theory,wolfe2019inflation,gisin2019entanglement,renou2019genuine,renou2021quantum, tavakoli2021bell}, motivating a number of novel experimental implementations ~\cite{carvacho2017experimental,saunders2017experimental,andreoli2017experimental,sun2019experimental,poderini2020experimental,chaves2018quantum,polino2019device,carvacho2019perspective,ringbauer2016experimental,agresti2021experimental}. Of particular, relevance is the bilocality scenario, providing the simplest network beyond the bipartite case and that for this reason has attract significant interest ~\cite{branciard2010characterizing,branciard2012bilocal,tavakoli2021bilocal,renou2021quantum,branciard2010characterizing,carvacho2017experimental,saunders2017experimental,andreoli2017experimental,sun2019experimental,poderini2020experimental}. Its causal structure is represented by the DAG in Fig.~\ref{fig:DAGnew}b, where two nodes (A,C) are connected with a central one (B), by means of two independent sources of correlations ($\Lambda_1$,$\Lambda_2$) \footnote{It is a particular case of the more general scheme, the star-shape network \cite{poderini2020experimental,tavakoli1,baumer2021demonstrating} (Fig.~\ref{fig:DAGnew}c). In this kind of network $n$-nodes are connected to a central one, that is $n=2$ for the bilocal case.}.

The classical description of the bilocal scenario is uniquely defined by the Markov condition~\cite{pearl2009causality}, which constrains the conditional probabilities of the measurement outcomes as
\begin{equation}
\begin{split}
p(a,b,c \vert x_A,x_B,x_C) = \sum_{\lambda_1,\lambda_2} p(\lambda_1)p(\lambda_2) \times\\\times p(x_1 \vert x_A,\lambda_A)p(x_B \vert x_B,\lambda_2)p(x_C \vert x_C,\lambda_1,\lambda_2),
\end{split}
\label{eq:DAGBell2}
\end{equation}
in which the independence of the sources of shared randomness implies the non-linear condition $p(\lambda_1,\lambda_2)=p(\lambda_1)p(\lambda_2)$. If one out of two possible dichotomic measurements is performed, that is,  $a,b,c \in \{0,1\}$ and $x_A, x_B, x_C \in \{0,1\}$, the classical description \eqref{eq:DAGBell2} implies that the observed correlations should respect a non-linear Bell inequality given by~\cite{branciard2010characterizing,branciard2012bilocal,tavakoli2014nonlocal,chaves2016polynomial,rosset2016nonlinear}
 \begin{equation}
\mathcal{B}=\sqrt{\vert I_1 \vert}+\sqrt{\vert I_2 \vert}\leq 1.
\label{eq:Bell2}
\end{equation}
where
\begin{equation}
\begin{split} 
I_1 &= \frac{1}{4}\sum_{x_A,x_B} \langle A^{x_A} B^{x_B} C^{0} \rangle, \\
 I_2 &= \frac{1}{4}\sum_{x_A,x_B} (-1)^{x_A+x_B} \langle A^{x_A} B^{x_B} C^{1} \rangle, \\
 \langle A^{x_A} B^{x_B} C^{x_C} \rangle &= \sum_{a,b,c} (-1)^{a+b+c} p(a,b,c \vert x_A,x_B,x_C) \;.
\end{split}
\label{eq:IBell2}
\end{equation}

In a quantum description of the bilocal experiment, the two sources are represented by bipartite quantum states $\rho_1$ and $\rho_2$ and the measurements are given by observables defined by operators $\hat{A}^{x_A}, \hat{B}^{x_B}, \hat{C}^{x_C}$ acting on their respective subsystem.
The corresponding probability distribution of measurement outcomes are given, according to the Born's rule, by $p(a,b,c\vert x_A,x_B,x_C) = \Tr \left[\left(\hat{A}^{x_A} \otimes\hat{B}^{x_B} \otimes \hat{C}^{x_C}\right) \left( \rho_{1} \otimes \rho_{2} \right)\right]$. As demonstrated by several works using photonic platforms~\cite{carvacho2017experimental,saunders2017experimental,andreoli2017experimental,sun2019experimental}, by properly choosing the quantum states and measurements, the bilocality inequality \eqref{eq:Bell2} can be violated, showing the incompatibility between classical causality and quantum predictions also in this new kind of causal network.
Remarkably, adopting separable measurements in the central node B it is also possible to violate such inequalities \cite{branciard2012bilocal,gisin2017all,andreoli2017maximal,andreoli2017experimental,poderini2020experimental}. In particular, when singlet quantum states, $\ket{\Psi^-_{1}}$ and $\ket{\Psi^-_{2}}$, are prepared by the two sources and distributed between the parties, a maximum violation of the bilocal Bell-like inequality can be obtained
$\mathcal{B}^{\mathrm{max}}_\mathrm{Q}=\sqrt{2} \approx 1.414$.
This value is achieved by considering two separable observables in the central node B, which measure the subsystems of the singlet state shared with A and C: $\hat{B}^{x_B} = \hat{B}_A^{x_B}\otimes\hat{B}_C^{x_B}$. Both observables can be taken as Pauli matrices $\sigma_z$ and $\sigma_x$. While each of the external nodes measures $(\sigma_x+\sigma_z)/\sqrt{2}$ and $(\sigma_x-\sigma_z)/\sqrt{2}$. 
This condition of separable measurements is particularly suitable to guarantee the scalability of the network, since it can be implemented using independent photonic platforms without the stringent requirements needed to perform entangled measurements. This is precisely our case, where we even adopt different quantum emitters of single photons, a SPDC source and a QD pumped by independent lasers working in continuous and pulsed mode, respectively. Importantly, the violation of the bilocality inequality \eqref{eq:Bell2} is possible even if the distribution $p(a,b,c\vert x_A,x_B,x_C)$ does not violate any standard 
Bell inequality. For instance, the choices of states and measurements described above  and that will be used in our experimental implementation, cannot violate the Clauser-Horne-Shimony-Holt (CHSH) inequality \cite{clauser1969proposed} between stations A and B,  given by
\begin{equation}
S_{\mathrm{AB}} = \sum_{x_A,x_B} (-1)^{x_Ax_B} \langle A^{x_A} B^{x_B} \rangle \leq 2, 
\label{eq:CHSH}
\end{equation}
and similarly cannot violate the corresponding CHSH inequality between stations B and C. 
That is, the non-classicality of the considered statistics truly requires the test of the underlying bilocality network to be detected.

\emph{Experimental apparatus --} In the following we discuss the photonic platform implementing bilocality network and achieving the violation of the inequality \eqref{eq:Bell2}. See Fig.~\ref{fig:DAGbilocale} for details.

One of the sources ($\rho_1$) of polarization-entangled photons is composed by a single GaAs QD in a matrix of Al$_{0.4}$Ga$_{0.6}$As. The quantum dots are fabricated with the Al droplet etching method and are placed between two asymmetric distributed Bragg reflectors, as detailed in \cite{BassoBasset2019}. Using a Weierstrass solid immersion lens and an aspheric lens for collection, the extraction efficiency is approximately $10\%$. The QD is optically pumped under resonant two-photon excitation \cite{Jayakumar2013} at $782.2$ nm with a laser repetition rate of 320 MHz. The rate of measured coincidence events at the output of the source 
is $13.7$ kHz. The QD is selected with a low fine-structure splitting (FSS)  of $0.8 \pm 0.5$ $\mu$eV to achieve a measured fidelity to a maximally entangled state of $0.929 \pm 0.004$, without the need for temporal or spectral filtering. The  CHSH parameter \eqref{eq:CHSH} measured 
is $S_{\mathrm{QD}} = 2.66 \pm 0.02$. Using a set of wave plates we maximize the fidelity of the photon pairs to the singlet state in polarization.

Regarding the SPDC source ($\rho_2$), entangled photon pairs in polarization are generated pumping a nonlinear ppKTP crystal placed into a Sagnac interferometer. The continuous-wave pump has a frequency of $405$ nm, while the signal-idler generation is degenerate at $810$ nm. The rate of measured coincidence events between the two output modes is $3$ kHz. The final state is properly tuned by a polarization controller in order to obtain singlet states in polarization, $\ket{\Psi^-}$, along the two output modes (1,2). The  CHSH parameter \eqref{eq:CHSH} measured at the output of the source, i.e. before distributing the photons, is  $S_{\mathrm{SPDC}}=2.727\pm0.007$ without accidental correction of the data. The estimated state fidelity with respect to the singlet state is $0.955\pm0.001$. In addition, the source is mounted in a compact and monolithic architecture (see Supplemental Material) providing high stability of the generated signal, as well as enabling the possibility to transfer and operate the source in different locations.

Using these sources of photonic entanglement, the implemented quantum bilocal network is reported in Fig.~\ref{fig:DAGbilocale}.
The source $\rho_{1}(\rho_{2})$ is shared between external A(C) and the central node B. The two sources are located near the corresponding peripheral nodes, i.e. source $\rho_{1}(\rho_{2})$ is in the same laboratory as the measurement station A(C). Two stations, B and C, are placed in two distinct laboratories inside the same building, and are connected though a $25$ m long single mode fiber. The other station A is placed in a different building $270$~m apart from the one of B and C. This node is connected to the central one (B) through a free-space channel stabilized by means of dedicated system using the feedback from an additional reference laser at $850$ nm and a couple of piezoelectric mirrors at the receiver to counter the effects of atmospheric turbolence and beam wander~\cite{basset2021quantum}, plus a piezoelectric mirror at the sender to remove thermal drifts in the pointing direction (see Supplemental Material). 
Stations C and B are also equipped with a GPS-locked oscillator, used to generate the common clock signal necessary for the synchronization of the detection events between the two buildings. The system also exploits the two-fold coincidences events between C and B to obtain the required accuracy.
In order to realize the optimal measurements, each station is equipped with standard setups for linear polarization measurement, consisting of a half wave plate (HWP) and polarizing beam splitter (PBS). Specifically, in the external nodes, A and C, the two basis required are realized by setting the HWP at $11.25^{\circ}(=\pi/16)$ and $33.75^{\circ}(=3\pi/16)$. 
Conversely, in the central node B, the measurements performed on the photons coming from the two quantum channels  are realized using two independent measurement setups. In both these setups, it is sufficient to set the HWP at angles $0^{\circ}$ and $22.5^{\circ}(=\pi/8)$.

\begin{figure}[ht!]
\centering
\includegraphics[width=0.99\columnwidth]{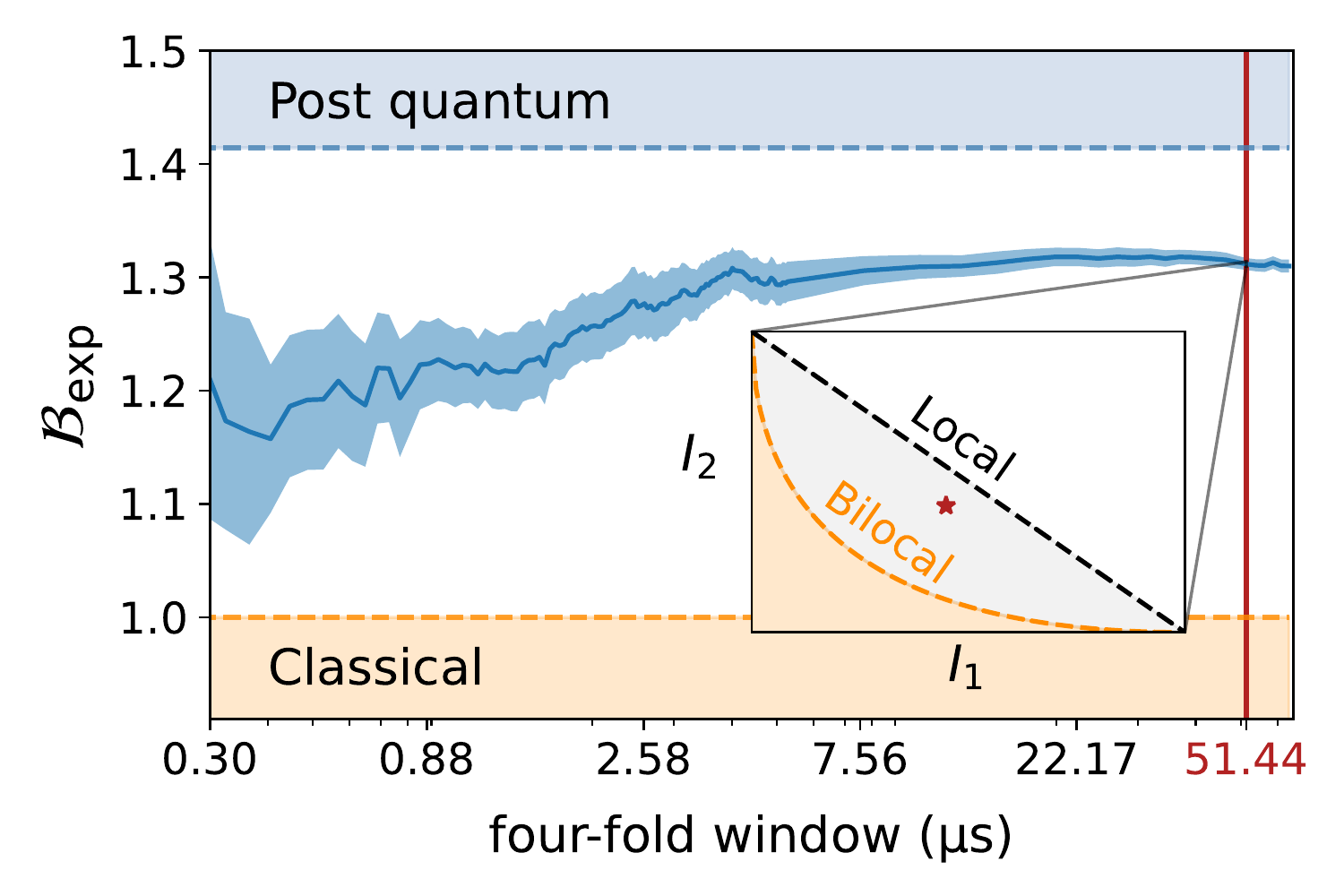}
\caption{ \textbf{Experimental results.} Quantum violation $\mathcal{B}_\mathrm{exp}$ of the Bell-like inequality for the bilocal scenario is shown as a function of the time window in which four-fold coincidence events are considered simultaneous. The dashed orange and blue lines represent the classical and the quantum bounds, $\mathcal{B}_\mathrm{cl} = 1$ and $\mathcal{B}_\mathrm{Q} = \sqrt{2}$ respectively, while the the solid blue line indicates the measured value of $\mathcal{B}_\mathrm{exp}$ with the corresponding error represented by the light blue area. The error is computed through repeated measurements over $\sim 25$ minutes. While the optimal value, in terms of $\sigma$-distance from the bound, is reached using a window of $51.4350 \mu$s, a significant violation can be obtained also considering much smaller windows, up to $283.5$ns.
In the inset, the point corresponding to the optimal window of $51.4350 \mu$s is
shown in the space of correlations $I_1, I_2$, as defined in Eq.~\eqref{eq:IBell2}, where the classical bound corresponding the the bilocal scenario is represented by the green area, while the gray area represents the classical correlations allowed by the relaxed scenario in which the assumption of independence of the source is lifted, corresponding to a tripartite Bell scenario.
}
\label{fig:results}
\end{figure}

\emph{Results -- }
Each observable of the tripartite system ABC requires the simultaneous detection of four-fold coincidence events. In order to achieve this, we first record two-fold events of subsystems AB and BC, independently. Then, four-fold events are recovered by filtering all the two-fold data inside a given time window. This procedure allows us to considerably reduce background noise while retaining an optimal rate for the four-fold events. The size of the window can be tuned, thus varying the statistics and the considered simultaneity of the events, similarly to \cite{poderini2020experimental}. The analysis of the violation of Eq.\eqref{eq:Bell2} is reported in Fig.~\ref{fig:results}, where optimal performances are reached for the window $51.435$ $\mu$s. Using this window, we obtain a mean value of $\mathcal{B}_{\mathrm{exp}}=1.312\pm0.0052$ averaged over $\sim25$ minutes. Further, two-fold events can be processed to extract the values of Bell parameters distributed along the quantum channels during the bilocality experiment. The analysis provides  CHSH violations of $S_{\mathrm{AB}}=2.484\pm0.018$ and $S_{\mathrm{BC}}=2.699\pm0.006$ for the QD and the SPDC source, respectively. 
The time trend of CHSH and bilocality violation of our network can be found in Supplemental Material. 
Notably, we obtain a violation of the classical limit also with windows in the range $300$--$800$~ns which, in principle, allows the events in nodes A and B to be recorded with space-like separations with respect to C, partially addressing the locality loophole of our implementation.

\emph{Conclusions -- }
A crucial requirement for the development of quantum communication networks is the ability to exploit and combine widely different technologies that are currently available, in a modular and reliable way. In this direction, we experimentally demonstrated the quantum violation of local causality in a hybrid tripartite quantum network. We violate a bilocality inequality, surpassing the classical bound by more than 60 standard deviations and thus prove the emergence of nonclassical correlations that could not be detected by standard Bell inequalities. Our network is composed of three nodes interconnected by fiber and free-space photonic links and two distinct sources of entangled photons. The two sources are based on significantly different technologies: a SPDC-based generation at $810$ nm pumped by $405$ nm cw-laser and a quantum dot emission at $781.2$--$783.2$ nm pumped in pulsed regime. Thus, our platform employs two intrinsically independent sources, a fundamental requirement for testing classical bounds in networks such as the bilocality scenario and others of increasing size and number of sources. Furthermore, we studied the violation of the Bell-like inequality for such scenario, tuning the time window in which four-fold coincidence events are considered simultaneous, including time windows in which the events in the distant stations can be considered space-like separated, an important ingredient for a fully loophole free demonstration. This work shows the reliability and versatility of the implemented platform, merging and interfacing technologically different solutions in the same network. The use of separable measurements allows interfacing sources of different natures, avoiding the drawbacks of optical Bell-state measurements requiring also the synchronization of the single-photon emission \cite{Huber2017}. Furthermore, our experiment has employed both free-space and fiber links in an urban environment, whose combined adoption represent a crucial requirement towards large scale networks. All these features demonstrate that our approach can be easily applied and scaled to any complex causal network, and can be used as a building block for future real-life quantum secure communication networks based on quantum nonlocality.

\emph{Acknowledgements --}
 This work was financially supported by the European Research Council (ERC) under the European Union’s Horizon 2020 Research and Innovation Programme (SPQRel, grant agreement no. 679183), and by MIUR (Ministero dell’Istruzione, dell’Università e della Ricerca) via project PRIN 2017 “Taming complexity via QUantum Strategies a Hybrid Integrated Photonic approach” (QUSHIP) Id. 2017SRNBRK. We acknowledge support from The John Templeton Foundation via the grant Q-CAUSAL No. 61084 and via The Quantum Information Structure of Spacetime (QISS) Project (qiss.fr) Grant Agreement No.  61466 (the opinions expressed in this publication are those of the authors and do not necessarily
reflect the views of The John Templeton Foundation). This project has received funding from the European Union’s Horizon 2020 Research and Innovation Program under Grant Agreement no. 899814 (Qurope). RC also acknowledges the Serrapilheira Institute (Grant No. Serra-1708-15763), the Brazilian National Council for Scientific and Technological Development (CNPq) via  the INCT-IQ and Grants No. 307172/2017-1 and No. 311375/2020-0.

\providecommand{\noopsort}[1]{}\providecommand{\singleletter}[1]{#1}%

\section{Supplementary Material} 

\section{1. Specifications of the experimental apparatus}

\subsection{A. Fiber and free-space quantum channels}

\begin{figure*}[t]
\centering
\includegraphics[width=\textwidth]{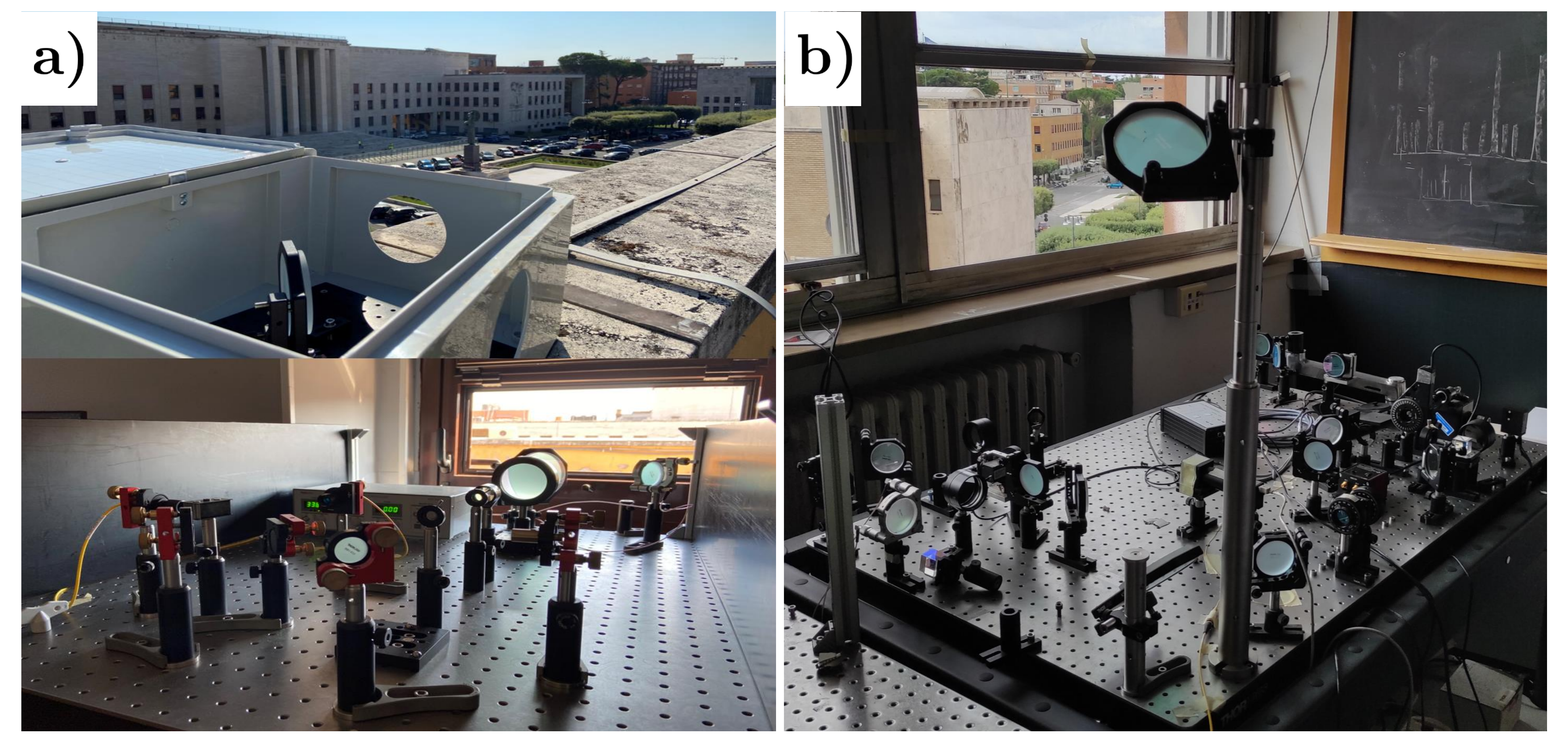}
\caption{(a) Transmitter and (b) receiver platforms adopted for the free-space optical link. The platforms are in an air-conditioned laboratories to mitigate the impact of environmental effects such as temperature variations. In transmitter side, a mirror in an external enclosure is placed on the terrace of the building (above) to compensate for the lack of direct line of sight between the two laboratories.}
\label{fig:airlink}
\end{figure*}

As described in the main text, two channels for quantum communication have been employed: a free-space channel ($\mathcal{C}_{\mathrm{AB}}$) and a fiber link ($\mathcal{C}_{\mathrm{BC}}$) connecting A to B and C to B, respectively. $\mathcal{C}_{\mathrm{BC}}$ connects two measurement stations in the same building using a $25$m-long single mode fiber, which has shown losses of $8\%$. Conversely, the implementation and the characterization of the free-space link $\mathcal{C}_{\mathrm{AB}}$ required more sophisticated considerations. When light propagates in free-space, it is affected by several phenomena. Natural divergence of Gaussian beams spreads light out of its collimation. Then, the beam wandering, due to interaction of light with air masses having random density, continuously changes the beam direction. Reducing these effects requires a trade-off in the choice of beam waist to use. The larger the beam waist, the higher the collimation. Conversely, a reduced spot size of the beam is less affected by beam wander oscillations. In our case we adopted a beam waist diameter of $22$ mm, thus providing a Rayleigh range of about $2$ km. This is done by a suitable telescope in the sender platform (Fig.\ref{fig:airlink}a). The resulting beam wander must be corrected in order to handle the received light and couple it into single mode fiber. In order to achieve this task, we employed an advanced stabilization system: the MRC-Laser Beam Stabilization (by MRC Systems GmbH). It consists of two fast steering mirrors (FSMs) connected to as many position detectors (PDs) and a control unit. This system allows a stabilization of the beam center at frequencies $<200$ Hz with accuracy $<0.1 \mu$m, i.e. suitable for single-mode coupling. The entire receiver platform is reported in the Supplementary Fig.\ref{fig:airlink}b. The losses of the free-space channel can be minimized by a suitable choice of the signal wavelength. A good compromise for free-space propagation is the infrared region around $800$ nm. In our case, we adopted a single-photon signal at $783.2$ nm and a further laser at $850$ nm, which acts as stabilization control. At these wavelengths, the losses measured along the link in free-space are about $15\%$. At the sender platform, both the light at $783.2$ nm and the laser at $850$ nm exit from the same single mode fiber, in order to experience the same perturbations during the air link propagation. Once arrived at the receiver platform, the light is reduced by a telescope similar to the one exploited for the sender platform. The 783.2 nm-signal is divided by the 800 nm-laser through a dichroic mirror (DM), and coupled into a single-mode fiber. This fiber is connected to a further optical table, where the signal is measured according to the protocol described in the main text. Instead the control laser is transmitted by the DM and sent to PDs, thus enabling the active stabilization of the photons affected by the free-space channel. \\
Additionally, to compensate for slow thermal drifts in the pointing direction of the sender platform and improve long-term stability, a mirror with a piezoelectric mount is inserted after the second lens of its telescope. The motion of the piezoelectric actuators is driven using a custom PID controller. The feedback is given by a compact CMOS camera placed at the receiver’s end which tracks the position of the 850 nm beam. The camera receives the light partially reflected by another DM placed after the second lens of the receiver's telescope, which transmits the signal at 783.2 nm.\\
While the experiment is conducted sending single photons at $783.2$ nm, the channel characterization and alignment is made by a simulation laser at the same wavelength. Using this laser, the losses were measured and the final losses budget is reported in the Supplementary Table \ref{tab:losses} for all channels used.   

\begin{table}[ht!] \begin{center} \begin{tabular}{|c|c|}
\hline
\textbf{free-space link} & \textbf{ Losses}\\
\hline
\hline

\begin{tabular}{c} Propagation on sending table \end{tabular}  & \begin{tabular}{c} $15\%$ \end{tabular} \\ \hline
\begin{tabular}{c} Air propagation \end{tabular}  & \begin{tabular}{c} $15\%$ \end{tabular} \\ \hline
\begin{tabular}{c} Propagation on receiving table \end{tabular}  & \begin{tabular}{c} $20\%$ \end{tabular} \\ \hline
\begin{tabular}{c} SMF coupling of stabilized signal \end{tabular} & \begin{tabular}{c} $50\%$ \end{tabular} \\ \hline
\end{tabular}
\caption{Main optical losses affecting the signal in the free-space quantum channel. SMF: single mode fiber.} \label{tab:losses}
\end{center} \end{table}


\subsection{B. Monolithic SPDC source}

\begin{figure}
\centering
\includegraphics[width=\columnwidth]{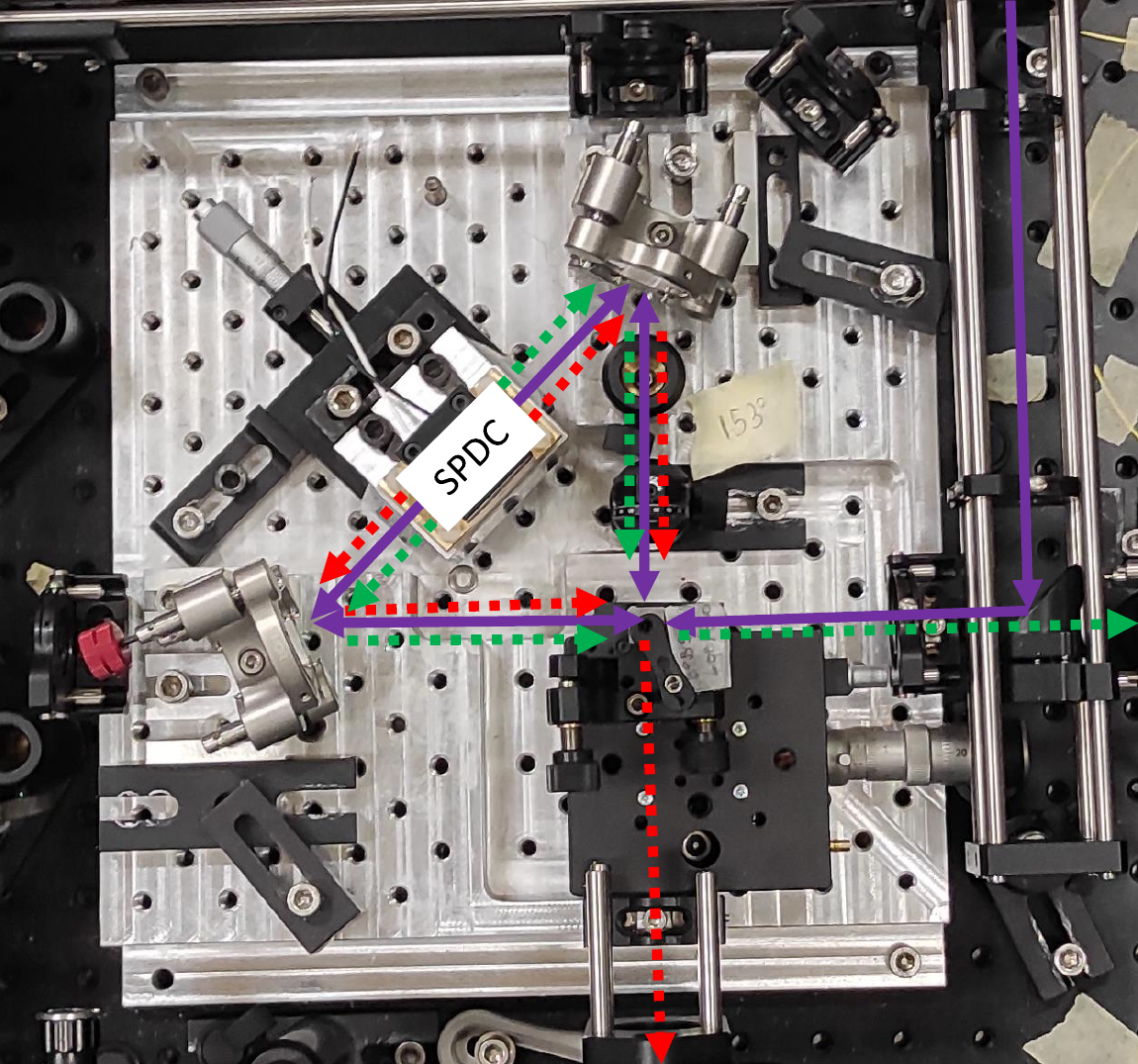}
\caption{Picture of the employed Sagnac source of polarization-entangled photon pairs. The pump laser at $405$ nm (purple line) generates pairs of photons (green and red dots) through SPDC process in a 2 cm ppKTP\footnote{Periodically poled potassium titanyl phosphate} non linear crystal. The recombination in the polarizing beam splitter allows the preparation of entangled states of light on the output modes of the interferometer. The monolithic architecture of the source provides high stability of the source performances. 
}
\label{fig:spdc}
\end{figure}

The realized source is based on a Sagnac interferometer that generates photon pairs entangled in polarization (see Fig.~\ref{fig:spdc}). It is mounted on a breadboard of reduced dimensions and makes use of a cage system in a compact and portable structure. This monolithic architecture demonstrated high stability of the produced signal over time, maintaining stable CHSH violation for over 24 hours. Furthermore, the possibility to be moved to different locations makes it interesting for further network investigation, as it is versatile to change structure geometry. For example, to realize node configurations where spatial loophole can be closed. 

\subsection{C. Synchronization between the three nodes}

During the measurement, the events registered by the single-photon detectors in the three nodes are sorted and timed by a TDC (Time-to-Digital Converter), with a resolution of $81$~ps, and filtered to select only two-fold coincidences between simultaneous detections coming from the same source. Since each of the three nodes features its own independent TDC, a common time reference is needed for the proper synchronization of the devices.

The solutions used in our work are different for the two channels, i.e. the fiber link and the free-space one. In particular, while for the former a common clock signal is directly shared between the two TDCs via coaxial cables, for the latter the larger distance that separates the stations makes this more problematic. We have solved it by resorting to two independent GPS-locked oscillators generating a reference signal, and then using the photon coincidences themselves to obtain the required sub-ns accuracy. In this way, we also demonstrate the feasibility of this last synchronization method, which can be effectively deployed in a wide variety of situations, where direct cabling is unavailable. In the following we give a brief description of the synchronization procedure.

\begin{figure}[tb!]
\centering
\includegraphics[width=0.9\columnwidth]{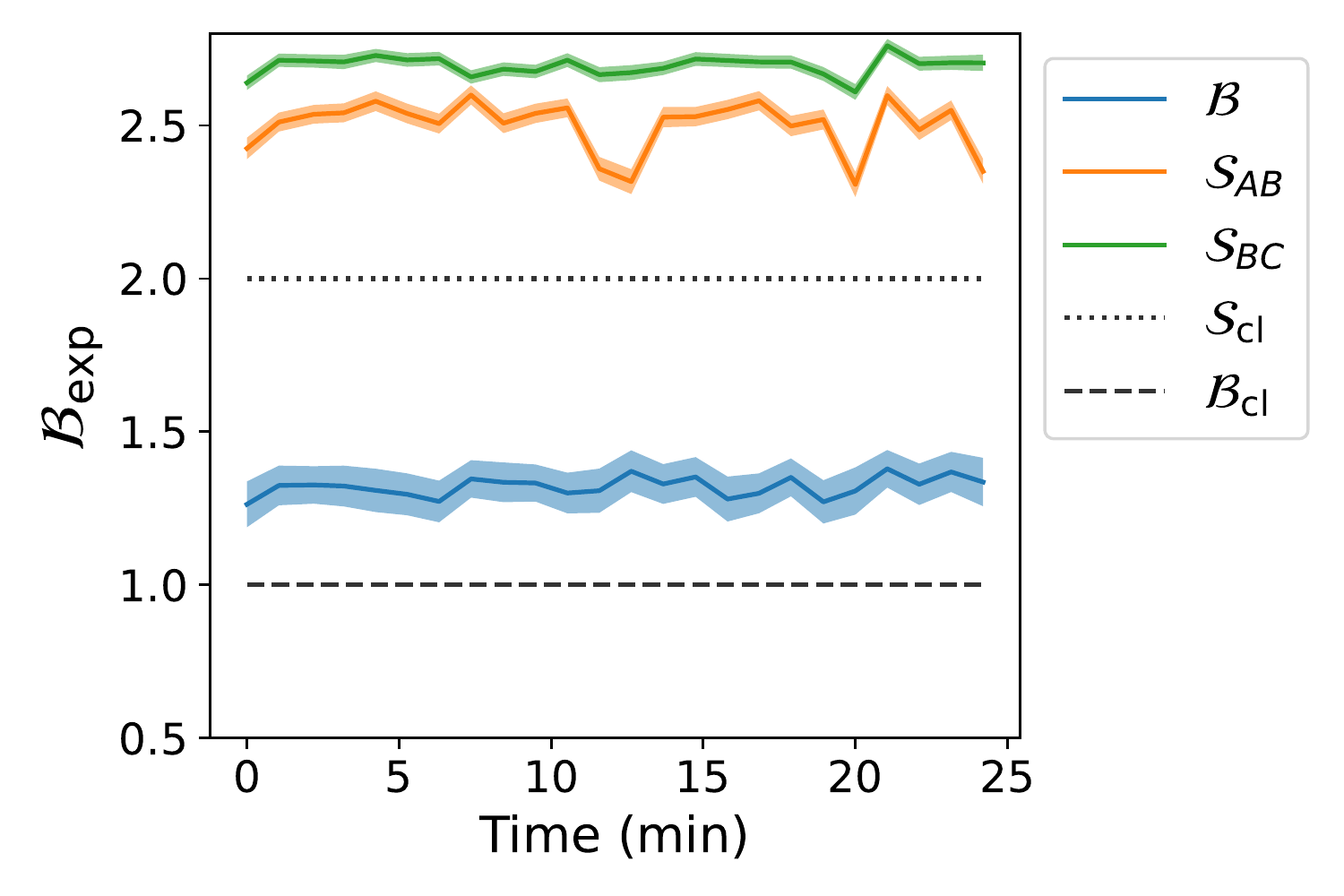}
\caption{Experimental violation of bilocality (blue curve) and individual CHSH violations (green and orange curves) are reported as a function of the time ($\sim25$ min of acquisition). Coloured shaded regions represent the corresponding standard deviations. Classical bounds are shown for comparison (dashed lines).}
\label{fig:biloc}
\end{figure}

The GPS-locked oscillators are located in stations A and B, and are configured to send two different TTL signals: a $10$~kHz square wave and a short pulse at $1$~Hz to their respective TDCs. The former signal is necessary to correct the drift of the internal TDC clock, while the latter represents the actual common clock. Node B shares these signals directly with C, which already gives an accurate synchronization between these two nodes. Node A instead relies on the GPS-locking between the signals in A and B which allows for a coarse synchronization up to $\approx 10$-$20$~ns.

The fine correction to the delays between node A, B and C are then performed by choosing two reference detectors, where two-fold coincidences are expected, and by looking at the highest peak in the histogram of the time differences between their detection events. The GPS synchronization described earlier allows us to efficiently perform this computation by restricting the histogram to a small region ($\pm 40$~ns) around the coarse value for the delay, which makes the procedure fast enough to be done in realtime while the measurement is running.

\section{2. Experimental violation of the bilocal scenario}

As reported in the main text, during the experiment on the tripartite system four-fold coincidences are employed to compute the bilocal parameter, while two-fold events to monitor the corresponding CHSH violations. Supplementary Figure \ref{fig:biloc} shows such parameters as a function of time and the stability of the network performances.


\end{document}